\documentclass{article}
\usepackage{kr}

\usepackage{times}
\usepackage{soul}
\usepackage{url}
\usepackage[hidelinks]{hyperref}
\usepackage[utf8]{inputenc}
\usepackage[small]{caption}
\usepackage{graphicx}
\usepackage{amsmath}
\usepackage{amsthm}
\usepackage{booktabs}
\usepackage{algorithm}
\usepackage{algorithmic}
\usepackage{booktabs}       %
\usepackage{enumitem}
\usepackage{cleveref}
\usepackage{microtype}      %
\usepackage{subfig}
\usepackage{arydshln}
\usepackage[textsize=tiny]{todonotes}
\newcommand{\solproj}[2]{\ensuremath{\mathsf{Sol}(#1)_{\downarrow{#2}}}}

\newcommand{\sol}[0]{\ensuremath{\mathsf{Sol}}}

\newcommand{\gnk}[0]{\ensuremath{\mathsf{Ganak}}}
\newcommand{\sstd}[0]{\ensuremath{\mathsf{SharpSAT}}-\ensuremath{\mathsf{TD}}}
\newcommand{\shst}[0]{\ensuremath{\mathsf{SharpSAT}}}
\newcommand{\dfour}[0]{\ensuremath{\mathsf{D4}}}
\newcommand{\arjun}[0]{\ensuremath{\mathsf{Arjun}}}

\newcommand{\addmc}[0]{\ensuremath{\mathsf{ADDMC}}}
\newcommand{\exmc}[0]{\ensuremath{\mathsf{ExactMC}}}
\newcommand{\apmc}[0]{\ensuremath{\mathsf{ApproxMC}}}

\newcommand{\gpmc}[0]{\ensuremath{\mathsf{GPMC}}}

\newcommand{\citebyname}[1]{\citeauthor{#1}~(\citeyear{#1})}
\newcommand{\mypar}[1]{\vspace{0.2cm} \noindent \textbf{#1}}
\newcommand{\mypari}[1]{\vspace{0.2cm} \noindent \textit{#1}}

\newcommand{\tot}[0]{2262}
\newcommand{\numdom}[0]{11}

\title{Model Counting in the Wild\thanks{  A preliminary version of this work appears at the 21st International Conference on Principles of Knowledge Representation and Reasoning, KR, 2024. The benchmarks and logfiles are available at \texttt{\color{blue} \url{ https://doi.org/10.5281/zenodo.13284882}}.}}

\author{
Arijit Shaw$^{1,2}$
\and
Kuldeep S. Meel$^3$%
\affiliations
$^1$Chennai Mathematical Institute, India\\
$^2$IAI, TCG CREST, Kolkata, India\\
$^3$University of Toronto, Canada%
}

\begin{document}

\maketitle

\begin{abstract}

  Model counting is a fundamental problem in automated reasoning with applications in probabilistic inference, network reliability, neural network verification, and more. Although model counting is computationally intractable from a theoretical perspective due to its \#P-completeness, the past decade has seen significant progress in developing state-of-the-art model counters to address scalability challenges.

  In this work, we conduct a rigorous assessment of the scalability of model counters in the \textit{wild}. To this end, we surveyed \numdom\ application domains and collected an aggregate of \tot\ benchmarks from these domains. We then evaluated six state-of-the-art model counters on these instances to assess scalability and runtime performance.

  Our empirical evaluation demonstrates that the performance of model counters varies significantly across different application domains, underscoring the need for careful selection by the end user. Additionally, we investigated the behavior of different counters with respect to two parameters suggested by the model counting community, finding only a weak correlation. Our analysis highlights the challenges and opportunities for portfolio-based approaches in model counting.

\end{abstract}

\section{Introduction}

Given a Boolean formula $F$ (often presented in conjunctive normal form), the problem of model counting is to compute the number of solutions of the formula. Model counting is a fundamental problem in computer science and has been studied by theoreticians and practitioners alike for the past four decades. From the perspective of theoreticians, model counting is a central problem in computational complexity: the seminal work of \citebyname{V79} established that the problem of model counting is \#P-complete, where \#P is the class of counting problems whose decision versions lie in NP. Toda's celebrated result~\cite{T89} showed that a single call to a \#P-oracle suffices to solve a problem in the entire polynomial hierarchy; formally, PH \(\subseteq \text{P}^{\text{\#P}}\). On the other hand, from practitioners' perspectives, model counting emerges as a central problem in a wide variety of domains such as quantitative software verification~\cite{TW21,GFS21}, probabilistic inference~\cite{Dar04}, network reliability~\cite{KM23}, cryptography~\cite{BZG20}, synthesis~\cite{GRM20}, product lines~\cite{SHTM+23,KKST+22}, neural network verification~\cite{BSSM+19}, and information flow~\cite{BAPP+16}. Consequently, despite theoretical hardness (in the worst case), there has been a demand for the development of algorithms and tools for model counting.

The earliest algorithmic approaches to model counting were pioneered in the early 2000s, combining advances in conflict-driven clause learning with knowledge compilation~\cite{Dar04,SBBKP04}. Subsequently, approaches in the early 2010s based on universal hashing and SAT solving were developed to obtain probably approximate counters~\cite{GSS06,CMV13}. Since then, there has been a significant surge of interest in the development of model counters. This development has led to substantial improvements in the runtime performance of state-of-the-art counters~\cite{Thu06,LM17,SRSM19,KJ21,LMY21}, best evidenced by the launch of the model counting competition in 2020~\cite{modelcounting2020}.

The model counting competition also allowed for the standardization of input and output formats, thereby making it easy to use different counters uniformly. The yearly competition also provides a snapshot of the performance of different counters. Given the annual snapshot of performance, one might be tempted to rely on the model counting competition to provide guidance on what counter to use in practice: for example, pick the winner of the latest model counting competition. Such a strategy is generally expected to fare well but may not be optimal since the objective of competitions is often to focus on benchmarks that are {\em difficult}. While selection focused on {\em difficult} benchmarks brings forth the weaknesses of state-of-the-art techniques, it may not guide the behavior of counters in the real world.

The primary objective of our investigation is the study of the scalability of model counters in {\em the wild}. To this end, we focused on the six state-of-the-art model counters, which have consistently performed well in model counting competitions over the past three years and rely on differing underlying techniques. As a next step, we constructed a benchmark suite comprising instances from  11 different application categories, including quantitative software verification, probabilistic inference, network reliability, cryptography, synthesis, product lines, neural network verification, and information flow. In total, our benchmark suite consists of 2262 instances. We then performed an extensive analysis of the performance of different counters in these instances. Furthermore, we sought to understand how the performance of counters varies with respect to features of input formulas.

Our experiments with different model counters on this set of benchmarks revealed the following:

\begin{enumerate}
    \item Among the individual solvers for non-projected instances, {\sstd} performed the best, solving 811 out of 1080 instances. For projected instances, {\apmc} achieved the highest performance, solving 1041 out of 1182 instances.
    \item Compilation-based, and hashing-based model counters excelled on different sets of benchmarks, often complementing each other. This complementary nature significantly improved the performance of the virtual best solver, which solved 2106 out of 2262 instances.
    \item The performance of compilation-based counters is correlated with treewidth, while the performance of hashing-based counters shows a weak correlation with independent support size.
\end{enumerate}

\mypari{Organization.} The structure of the paper is as follows. First, we introduce notation and preliminaries in \Cref{sec:notation}. Next, we discuss various algorithms for model counting in \Cref{sec:modelcounting} and describe the applications of model counting in \Cref{sec:benchmarks}. Our experimental results are presented in \Cref{sec:results}. Finally, we conclude in \Cref{sec:concl}.

\section{Notations and Preliminaries}
\label{sec:notation}
Let $X$ be the set of Boolean variables, and let $F$ be a Boolean formula in Conjunctive Normal Form (CNF) defined over variables in $X$. An assignment $\sigma: X \mapsto \{0,1\}$ is called a satisfying assignment or a solution if $\sigma$ makes $F$ evaluate to True. Given a set of projection variables $P \subseteq X$, a projection of assignment $\sigma$ to the set $P$ is the subset of assignments only to the variables of $P$.

\paragraph*{Model Counting.}
Let $\sol(F)$ denote the number of solutions of a given formula $F$. The model counting problem is determining $|\sol(F)|$.
\textit{An exact model counter} takes in formula $F$, and  returns $|\sol(F)|$. \textit{An  approximate model counter} takes in a formula $F$, tolerance parameter $\varepsilon$, confidence parameter $\delta$ and returns $c$ such that
$\Pr\left[\frac{|\sol(F)|}{1+\varepsilon} \leq c \leq (1+\varepsilon) |\sol(F)| \right] \geq 1-\delta$.

\paragraph*{Projected Model Counting.}
Let $\solproj{F}{S}$ denote the set of projected assignments satisfying the given formula $F$ and a projection set $S$. The problem of \textit{projected model counting} is to compute $|\solproj{F}{S}|$.
\textit{An exact projected model counter} takes in formula $F$, and  returns $|\solproj{F}{S}|$. An  \emph{approximate projected model counter} takes in a formula $F$, projection set $S$, parameters $\varepsilon$, and $\delta$, and returns $c$ such that $\Pr\left[\frac{|\solproj{F}{S}|}{1+\varepsilon} \leq c \leq (1+\varepsilon) |\solproj{F}{S}| \right] \geq 1-\delta$.

To differentiate between model counting and projected model counting, we use the term \textit{non-projected model counting} to refer to model counting without projection.

\paragraph{Independent Support.} For a given assignment $\sigma$ over $X$ and a subset of variables $S \subseteq X$, let $\sigma_{\downarrow S}$ represent the assignment of variables restricted to $S$. Given a Boolean formula $F$ over the set of variables $X$ and a projection set $S \subseteq X$, a subset of variables $\mathcal{I}$ such that $\mathcal{I} \subseteq S$ is called independent support (or simply \textit{support}) of $S$ if $\forall \sigma_1, \sigma_2 \in \sol(F), \sigma_{1\downarrow\mathcal{I}} = \sigma_{2\downarrow\mathcal{I}} \implies \sigma_{1\downarrow S} = \sigma_{2\downarrow S}$.
Several preprocessing techniques for model counting have been proposed, which compute a small independent support for the input formula and simplify the formula based on that support~\cite{LLM16,SM19}.

\paragraph{Treewidth.}
Treewidth is a measure of how {\em tree-like} a graph is. A tree decomposition of a graph $G = (V, E)$ is a pair $(T, \{B_i\}_{i \in I})$ where $T$ is a tree with nodes indexed by $I$, and  $\{B_i\}_{i \in I}$ are subsets of $V$ (bags) such that:
\begin{enumerate}
  \item Every vertex $v \in V$ is in at least one bag $B_i$.
  \item For every edge $(u, v)$, there is a bag $B_i$ with $u, v \in B_i$.
  \item For each vertex $v$, the bags containing $v$ form a connected subtree of $T$.
\end{enumerate}

The width of a tree decomposition $(T, \{B_i\}_{i \in I})$ is $ \max_{i \in I} (|B_i| - 1) $. The treewidth of $G$, denoted $\text{tw}(G)$, is the minimum width among all tree decompositions of $G$:
$ \text{tw}(G) = \min_{(T, \{B_i\})} \max_{i \in I} (|B_i| - 1) $.

\section{The Landscape of Model Counting}
\label{sec:modelcounting}

Significant progress has been made in developing efficient algorithms for model counting. In this section, we provide a brief overview of the different approaches.

\begin{enumerate}

      \item \textbf{Compilation-based Exact Model Counters.} The remarkable success of SAT solvers has motivated researchers to develop model counters that leverage the search techniques employed by these solvers. \citebyname{Dar04} introduced the use of deterministic decomposable negation normal form (d-DNNF) to efficiently obtain the model count in a model counter named $\mathsf{c2d}$. During the search procedure of  DPLL, the solver may encounter sub-formulas that have already been seen in a prior branch of the tree. To avoid redundant computations, it is essential to recognize such sub-formulas and reuse their model counts efficiently. To address this challenge, researchers have introduced the concept of component caching, which has led to the development of highly efficient model counters like $\mathsf{Cachet}$~\cite{SBBKP04} and {\shst}~\cite{Thu06}. \citebyname{LM17} further improved this approach by utilizing dynamic decomposition techniques to enhance the efficiency of d-DNNF-based techniques and designed the {\dfour} model counter. Subsequently, heuristics were integrated into component caching in the probabilistic model counter {\gnk}~\cite{SRSM19}. \citebyname{KJ21} combined the concept of tree decomposition was combined with component caching in {\sstd}. {\gpmc} uses a similar strategy as $\mathsf{SharpSAT}$, but optimizes it for projected model counting. \citebyname{LMY21} introduced a generalization of d-DNNF known as Constrained Conjunction and Decision Diagram (CCDD), which was implemented in the model counter {\exmc}.
            While all the aforementioned techniques employ a search-based top-down compilation approach, \citebyname{DPV20} utilized algebraic decision diagram (ADD) based bottom-up compilation methods to design the model counter {\addmc}, which also performs effectively due to its early-projection techniques.

      \item \textbf{Hashing-based Approximate counter.}
            Over the past decade, there has been the development of scalable approximate model counters that rely on XOR-based pairwise independent functions to divide the solution space into smaller parts and then invoke state-of-the-art SAT solvers to enumerate models in a randomly chosen cell to accurately estimate the model count~\cite{CMV13,CMV16,SM19,SGM20}. The state-of-the-art hashing-based counter, {\apmc}, has shown to work well in practice. It also supports preprocessing techniques based on independent set detection and scales better when {\arjun}~\cite{SM22} provides a small independent set.

\end{enumerate}

In the context of this survey, we focus on the six state-of-the-art model counters that have performed well in model counting competitions over the past year. The first five are top-down compilation techniques, with different technical improvements on top of the algorithm.

\begin{enumerate}
      \item {\sstd}: Developed on top of {\shst}, combined with a tree-decomposition-based heuristics.
      \item {\gnk}: Developed on top of {\shst}, enhanced with probabilistic component caching.
      \item {\dfour}: A Decision-DNNF compilation based on dynamic decomposition.
      \item {\gpmc}: Another top-down compilation-based counter, which uses optimizations for projected counting.
      \item {\exmc}: Another top-down compilation-based model counter using CCDD for compilation.
      \item {\apmc}: A hashing-based approximate model counter.
\end{enumerate}

Among these counters, {\exmc} and {\sstd} solve the problem of only non-projected model counting. The remaining solvers can solve the problems of both projected and non-projected model counting. We compare the performance of all the solvers in the categories in which they can solve the problem.

\section{Benchmarks}\label{sec:benchmarks}

We selected a large set of benchmarks from various practical domains to evaluate the model counters. Below is a list of these domains:

\begin{enumerate}

  \item \textbf{Software Verification.} In software verification, some quantitative problems are solved by reducing the problems to model counting. Here are two such problems:
        \begin{enumerate}
          \item \textit{Reliability Estimation.} When the functional correctness of a program cannot be established, a potential approach to assess the software's reliability is to quantify it as the ratio of failing program runs to all terminating runs. \citebyname{TW21} reduced this approach to model counting, where the model count corresponds to the number of inputs that trigger or bypass assertions or assumptions.

          \item \textit{Robust Reachability.} Determining the extent to which a bug can be replicated is frequently relevant. \citebyname{GFS21} addressed this issue by employing the formalism of robust reachability. They also introduced the concept of quantitative robust reachability, which seeks to identify a controlled input that maximizes the number of uncontrolled inputs capable of reaching the intended target. Model counting can be used to lower bound the runtime cost by the cost of determining the number of uncontrolled inputs satisfying a path constraint for a given controlled input.

        \end{enumerate}

  \item \textbf{Probabilistic Inference.} Model counting is used to solve the problem of probabilistic inference. \citebyname{SBK05} encoded the inference problem on Boolean Bayesian networks as a model counting problem.

  \item \textbf{Network Reliability.} For critical infrastructure like power transmission grids, it is important to know the reliability of the infrastructure. \citebyname{KM23} encoded the problem of network reliability as a weighted model counting problem. They then used chain formulas \cite{CFMV15} to encode the problem as unweighted model counting problems. These benchmarks encode power transmission grids from different cities and states. The number of solutions to these formulas relates to the network reliability.

  \item \textbf{Cryptography.} Certain problems in cryptography can also be tackled with model counting. \citebyname{BZG20} used model counting to automate the development of chosen-ciphertext attacks. %

  \item \textbf{Synthesis.} Given the specification of a function or program, the task of synthesis is to generate the function or program.
        \begin{enumerate}
          \item \textit{Program and Function Synthesis.} Some algorithms for synthesis \cite{GRM20} use model counts in certain parts. These benchmarks consist of instances where the specifications of the functions like arithmetic, disjunctive instances etc.
          \item \textit{Synthesis for Control Improvisation.} The control improvisation (CI) framework helps synthesize randomized systems with strict and flexible constraints. \citebyname{GVF22} includes quantitative constraints on expected costs and randomness constraints for diversity based on labels and encodes the problem as a model counting problem.
        \end{enumerate}

        \begin{table}[!btp]
          \centering
          \begin{tabular}{@{}lrr@{}}
            \toprule
            benchmark               & Treewidth & Support Size \\ \midrule
            Robust Reachability     & 9.13      & 1565.38      \\
            Bayes Net               & 15.52     & 2410.69      \\
            Industrial Config       & 18.29     & 749.45       \\
            Linux Config            & 21.50     & 799.32       \\
            Network Reliability     & 34.33     & 941.58       \\
            Control Improvisation   & 35.00     & 58.27        \\
            Cryptographic           & 41.49     & 292.97       \\
            Software Reliability    & 55.49     & 1111.65      \\
            Information Flow        & 66.84     & 10268.38     \\
            Functional Synthesis    & 83.57     & 681.82       \\
            Neural Net Verification & N.A.      & 71.26        \\ \bottomrule
          \end{tabular}
          \caption{Average Treewidth and Independent Support Size.}
          \label{tab:tw-ind}
        \end{table}

  \item \textbf{Feature Counting.} Product lines efficiently manage groups of products sharing a core set of features. Determining the number of valid configurations is often a crucial task.
        \begin{enumerate}
          \item \textit{Industrial Product Lines.} Product lines are commonly employed to handle families of products sharing a core set of features, with feature models serving as a standard to define valid feature combinations. However, not all feature configurations are permissible. These models facilitate standardized analyses of the system's variability, and many of these analyses require calculating the number of valid configurations. \citebyname{SHTM+23} surveyed these problems as a model counting problem.
          \item \textit{Configuration Spaces of Software Systems.} \citebyname{KKST+22} studied the problem of feature modeling, which helps systematically model features and dependencies in software systems. The authors encode feature models into propositional formulas, where the number of solutions of the formula corresponds to the number of possible features in a software system.

        \end{enumerate}

        \begin{figure*}[t]
          \begin{center}
            \subfloat[Non-projected instances]{\includegraphics[width=0.7\linewidth]{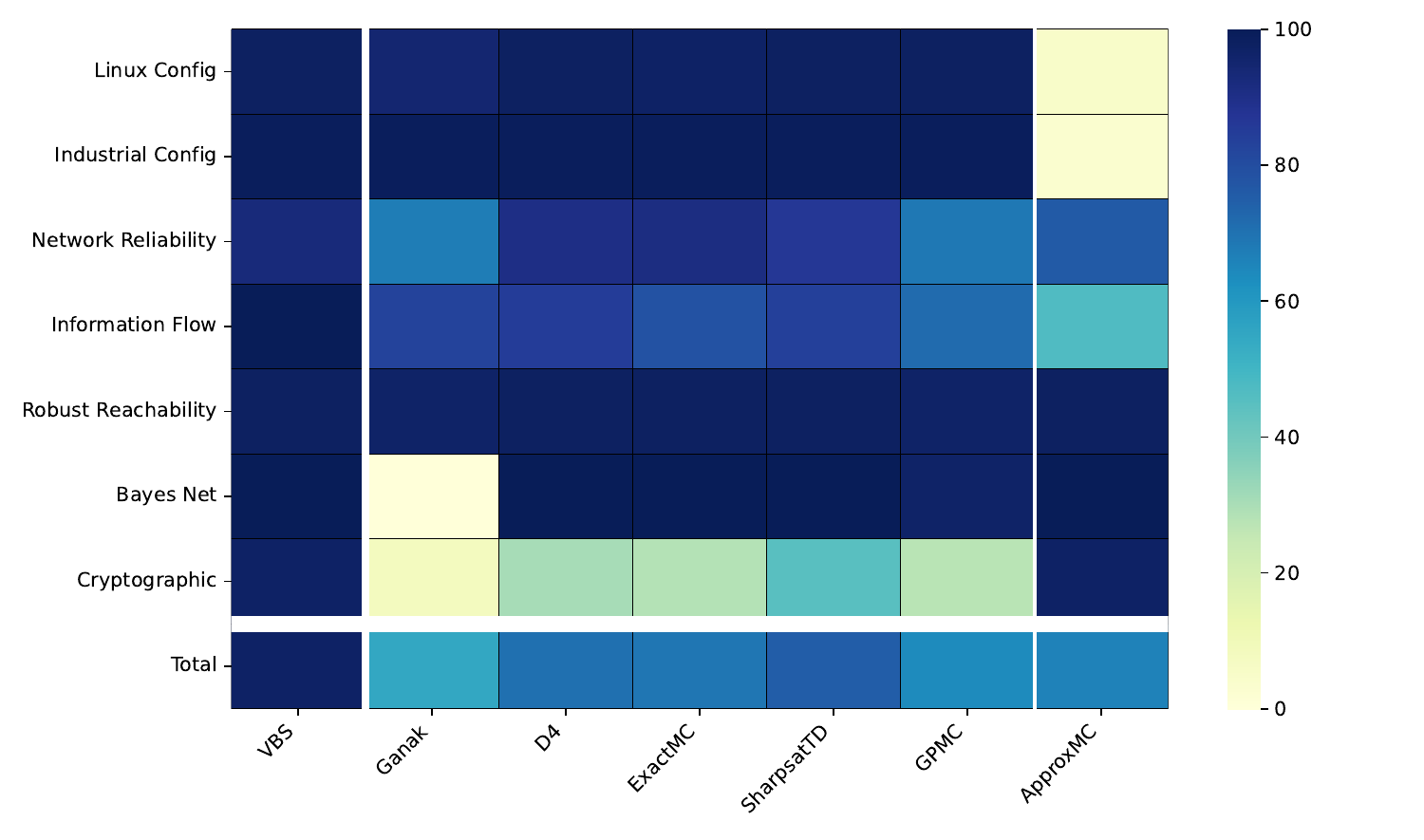}}
            \\
            \noindent
            \subfloat[Projected instances]{\includegraphics[width=0.75\linewidth]{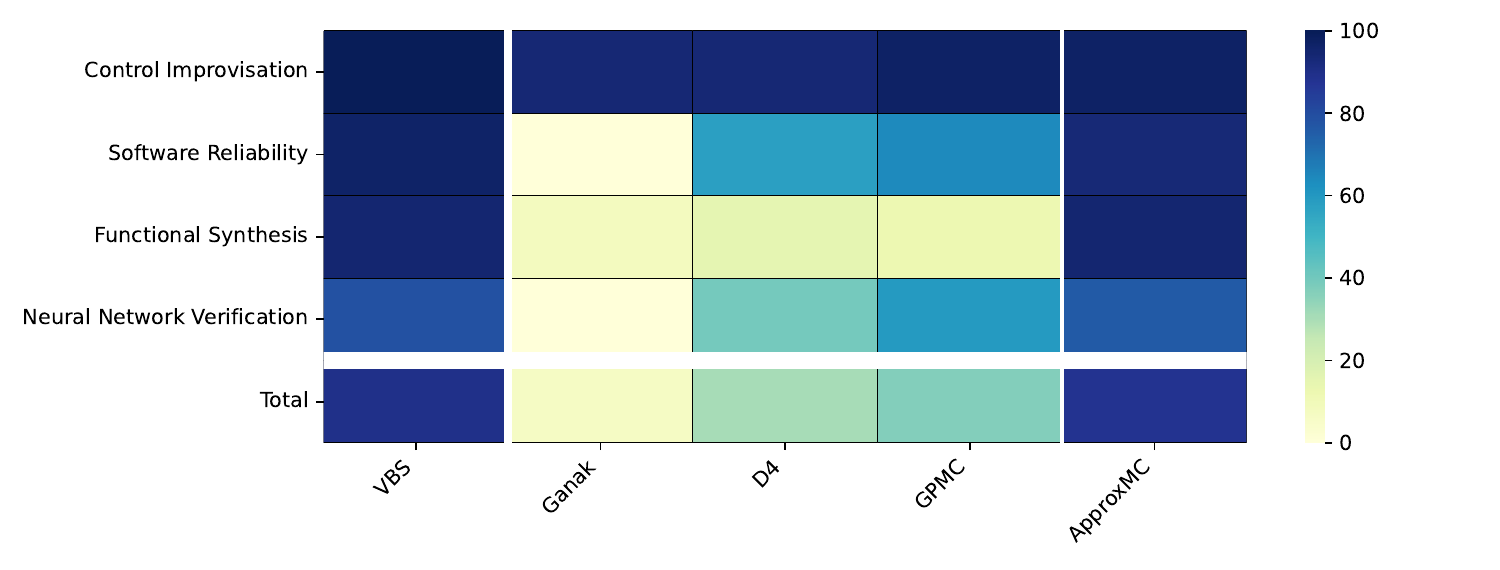}}
          \end{center}
          \caption{Heatmap of the percentage of instances solved by model counters for }
          \label{fig:heatmaps}
        \end{figure*}

  \item \textbf{Quantitative Verification of Neural Networks.} An intriguing aspect of neural network verification is assessing the frequency with which a specific property is valid. The NPAQ (Neural Property Approximate Quantifier) framework, introduced by \citebyname{BSSM+19}, facilitates the evaluation of various properties in binarized neural networks. The benchmarks test the following properties on the MNIST and UCI datasets: \textit{fairness}, which encodes bias towards marital status, race, or sex; \textit{robustness}, which measures the impact of 2-3-bit adversarial input perturbations; and \textit{Trojan attacks}, which account for the number of inputs with a trojan pattern. These benchmarks were initially encoded as pseudo-Boolean constraints and later converted to CNFs, thereby encoding many arithmetic circuits.

  \item \textbf{Quantitative Information Flow.} Information leaks in modern software systems are an important problem. \citebyname{BAPP+16} introduced an analysis method that estimates both minimum and maximum leak amounts, even when some paths aren't fully explored. This method was added to KLEE to analyze information leaks in C programs.

\end{enumerate}

Among these benchmark sets, functional synthesis, reliability estimation, control improvisation, and neural network verification benchmarks consist of projected counting instances, while the remaining are non-projected model counting problems.

\begin{table*}[!tbp]

  \centering
  \renewcommand{\arraystretch}{1.2}
  \begin{tabular}{@{}l|r|r|rrr|r@{}}
    \toprule
                                & Total & VBS  & Ganak & D4  & GPMC        & ApproxMC      \\ \midrule
    Control Improvisation       & 33    & 33   & 31    & 31  & \textbf{32} & \textbf{32}   \\ \hdashline{}
    Software Reliability        & 123   & 119  & 0     & 71  & 79          & \textbf{115}  \\
    Control Improvisation       & 33    & 33   & 31    & 31  & \textbf{32} & \textbf{32}   \\
    Functional Synthesis        & 609   & 577  & 50    & 94  & 74          & \textbf{577}  \\
    Neural Network Verification & 417   & 329  & 0     & 165 & 249         & \textbf{317}  \\ \midrule
    Total                       & 1182  & 1058 & 81    & 361 & 434         & \textbf{1041} \\ \bottomrule
  \end{tabular}
  \caption{Number of instances solved by different model counters on projected instances.}
  \label{tab:numsolved-proj}
  \vspace*{0.5cm}

  \begin{tabular}{l|r|r|rrrrr|r}
    \toprule
                        & Total & VBS   & Ganak        & D4           & ExactMC      & SharpsatTD   & GPMC         & ApproxMC     \\ \midrule

    Linux Config        & 135   & 130   & 126          & \textbf{130} & 129          & \textbf{130} & \textbf{130} & 7            \\
    Industrial Config   & 128   & {126} & \textbf{126} & \textbf{126} & \textbf{126} & \textbf{126} & \textbf{126} & 4            \\
    Network Reliability & 256   & 177   & 129          & 173          & \textbf{174} & 165          & 131          & 145          \\
    Information Flow    & 117   & 106   & 88           & \textbf{90}  & 83           & 89           & 76           & 50           \\  \hdashline{}
    Robust Reachability & 93    & 91    & 90           & \textbf{91}  & \textbf{91}  & \textbf{91}  & 90           & \textbf{91}  \\
    Bayes Net           & 29    & 29    & 0            & \textbf{29}  & \textbf{29}  & \textbf{29}  & 28           & \textbf{29}  \\ \hdashline{}
    Cryptographic       & 411   & 389   & 34           & 123          & 113          & 181          & 110          & \textbf{389} \\ \midrule
    Total               & 1169  & 1048  & 593          & 762          & 745          & \textbf{811} & 691          & 715          \\\bottomrule
  \end{tabular}
  \caption{Number of instances solved by different model counters on non-projected instances.}
  \label{tab:numsolved-unproj}
\end{table*}

\begin{figure*}[t]
  \begin{center}
    \subfloat[Cryptographic Benchmarks]{\includegraphics[width=0.45\linewidth]{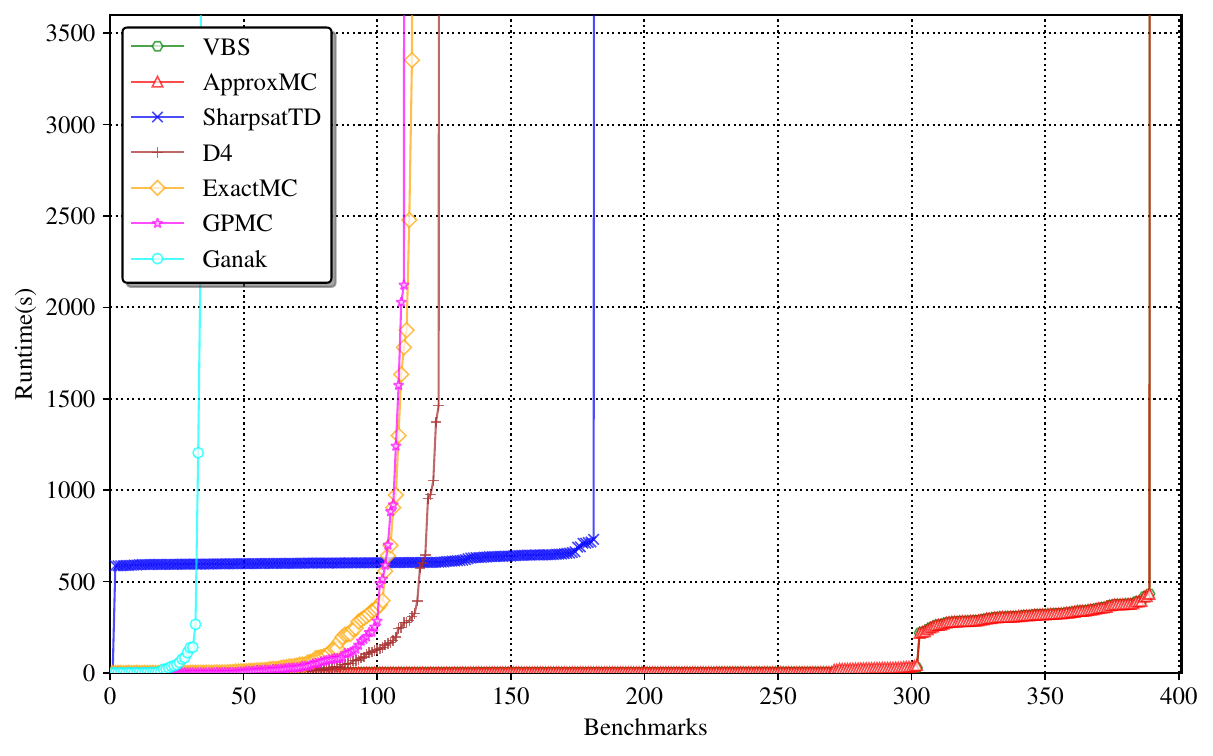}}
    \subfloat[Neural Net Verification Benchmarks]{\includegraphics[width=0.45\linewidth]{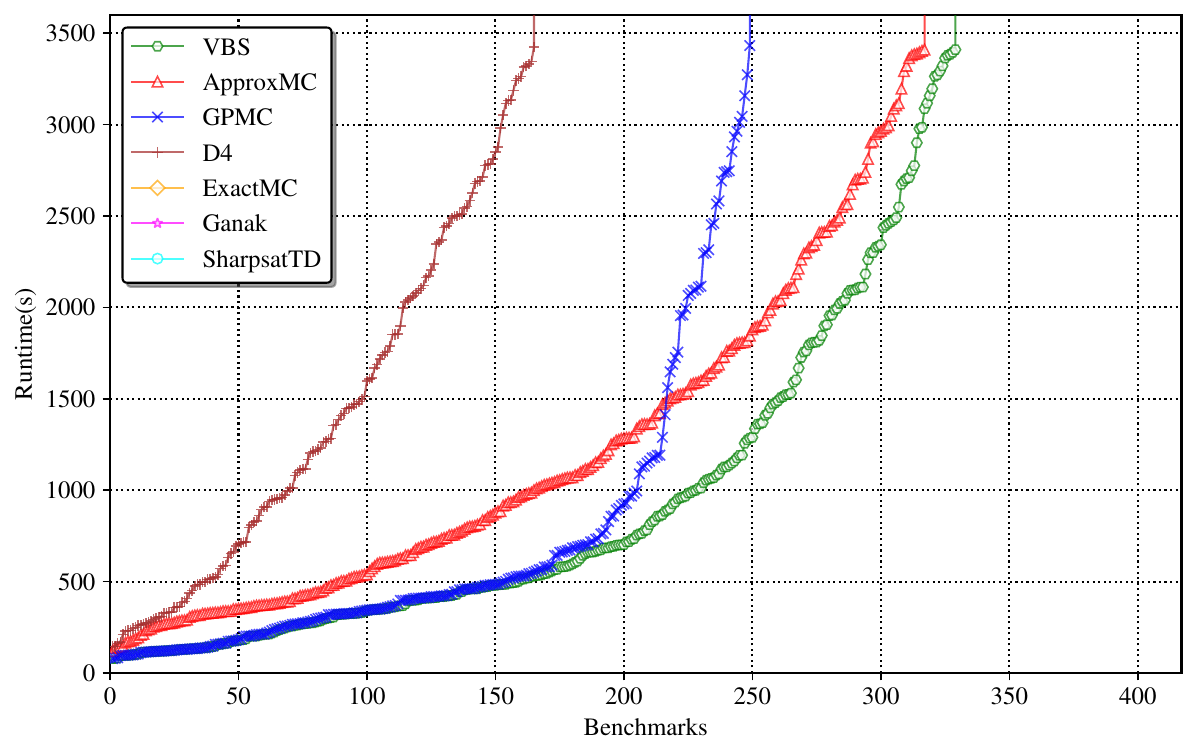}}
    \\
    \noindent
    \subfloat[Software Reliability Benchmarks]{\includegraphics[width=0.45\linewidth]{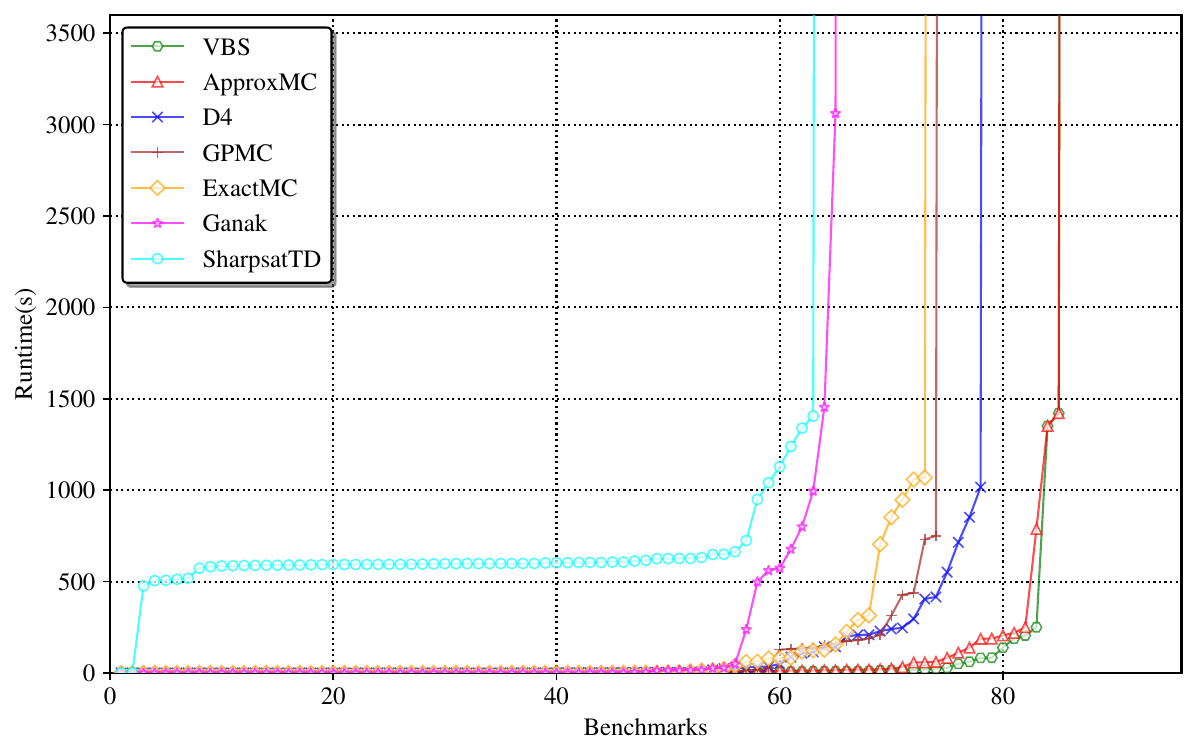}}
    \subfloat[Information Flow Benchmarks]{\includegraphics[width=0.45\linewidth]{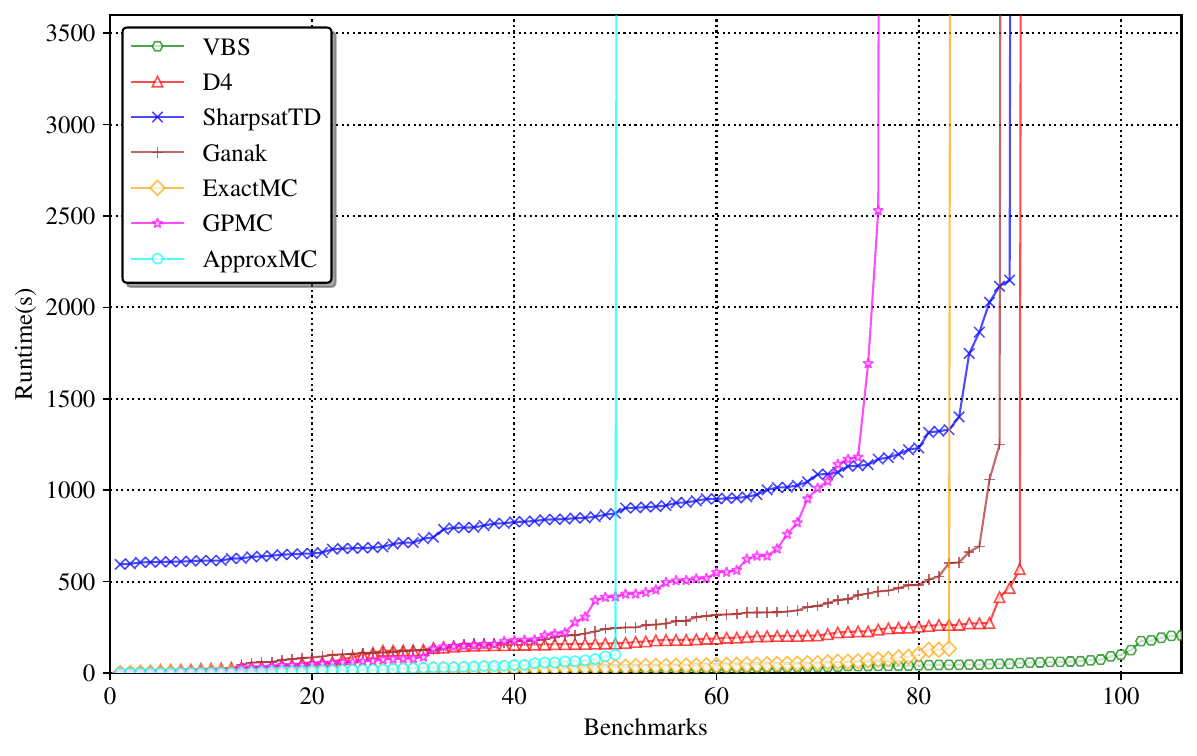}} %

  \end{center}
  \caption{Cactus plot of the numbers of benchmarks solved by model counters on different benchmark sets. }
  \label{fig:cactuss}
\end{figure*}

\section{Experimental Evaluation} \label{sec:results}

To evaluate the performance and effectiveness of the various tools discussed in \Cref{sec:modelcounting}, we conducted the following experiments.

\mypari{Experimental Setup.} The experiments were carried out on a supercomputing cluster equipped with AMD EPYC 7713 CPUs. Each experiment involved running a tool on a specific benchmark using a single core with a memory limit of 16 GB. For approximate counters, we set $\varepsilon = 0.8$ and $\delta = 0.2$. We adhered to the competition standard timeout of 3600 seconds. Initial experiments with a higher timeout showed a minimal increase in the number of instances solved. For the experiments, we used the versions of the counters submitted to the Model Counting Competition 2023.

\mypari{Virtual Best Solver.} We included the results of a Virtual Best Solver (VBS) in our comparisons. A VBS is a hypothetical solver that performs well and is the best method for each benchmark. If solvers $s_1, \dots s_n$ solve a problem in time $t_1, \dots t_n$ seconds, then VBS solves it in $\min(t_1, \dots, t_n)$ sec.

\mypari{Correctness.} The correctness of the model counters is well-established; in all model counting competitions, the counters consistently produce correct counts. Additionally, approximate model counters provide counts with deficient error. Therefore, we assume the counters are correct and do not focus on this aspect.

\noindent
Since not all model counters can handle projected model counting, we analyzed projected and non-projected instances separately.

\noindent
In this work, we sought to answer the following research questions:

\begin{enumerate}[font=\bfseries RQ, leftmargin=1cm]
  \item How do the overall performances of different model counters compare, and how do they vary across different sets of benchmarks?
  \item How do benchmark parameters relate to the performance of various solvers?
\end{enumerate}

\begin{table*}[!tbp]
  \renewcommand{\arraystretch}{1.2}
  \centering

\end{table*}

\mypar{Summary of Results.} The highest number of non-projected instances solved by a single solver was achieved by {\sstd}, which successfully solved 811 out of 1080 instances. For projected instances, {\apmc} demonstrated the best performance, solving 1041 out of 1182 instances. Compilation-based and hashing-based model counters excelled in solving different sets of benchmarks, and their performances were often complementary. This complementary performance resulted in a much better performance of the VBS, which solved 2106 out of 2262 instances. Treewidth correlates with the performance of compilation-based counters, while independent support size correlates weakly with the hashing-based counter.

\begin{figure*}[t]
  \begin{center}
    \subfloat[Network Reachability Benchmarks]{\includegraphics[width=0.45\linewidth]{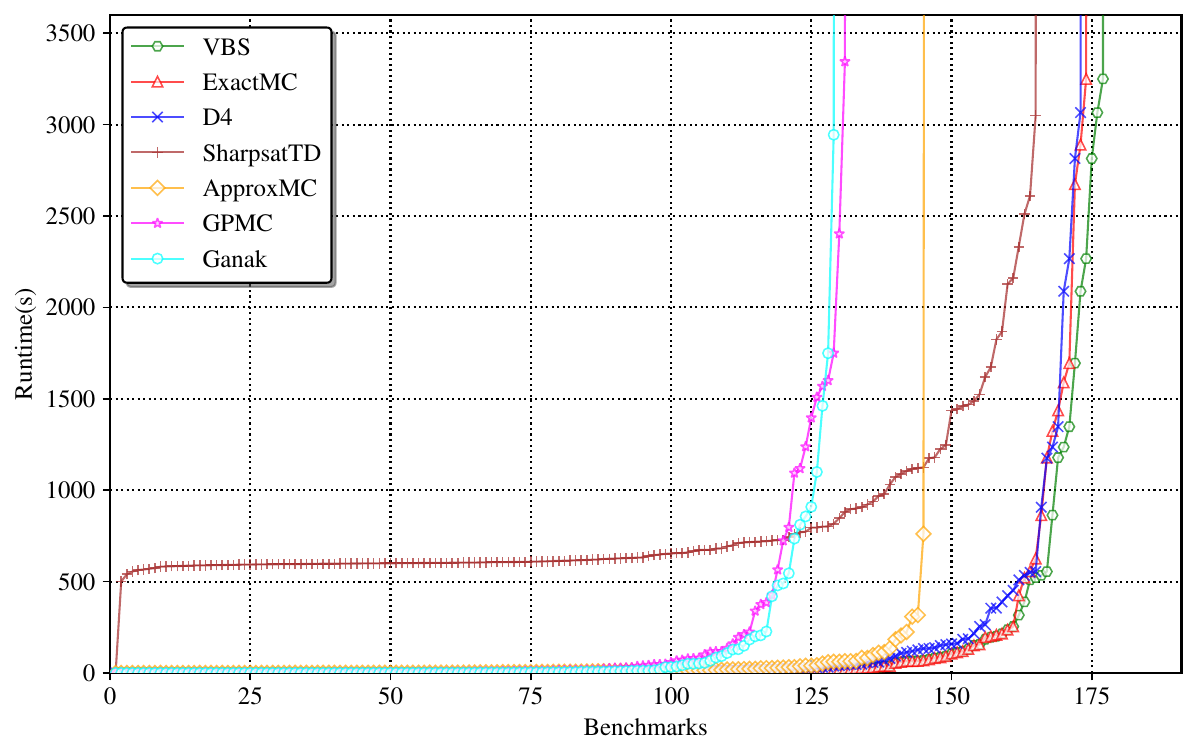}}
    \subfloat[Functional Synthesis Benchmarks]{\includegraphics[width=0.45\linewidth]{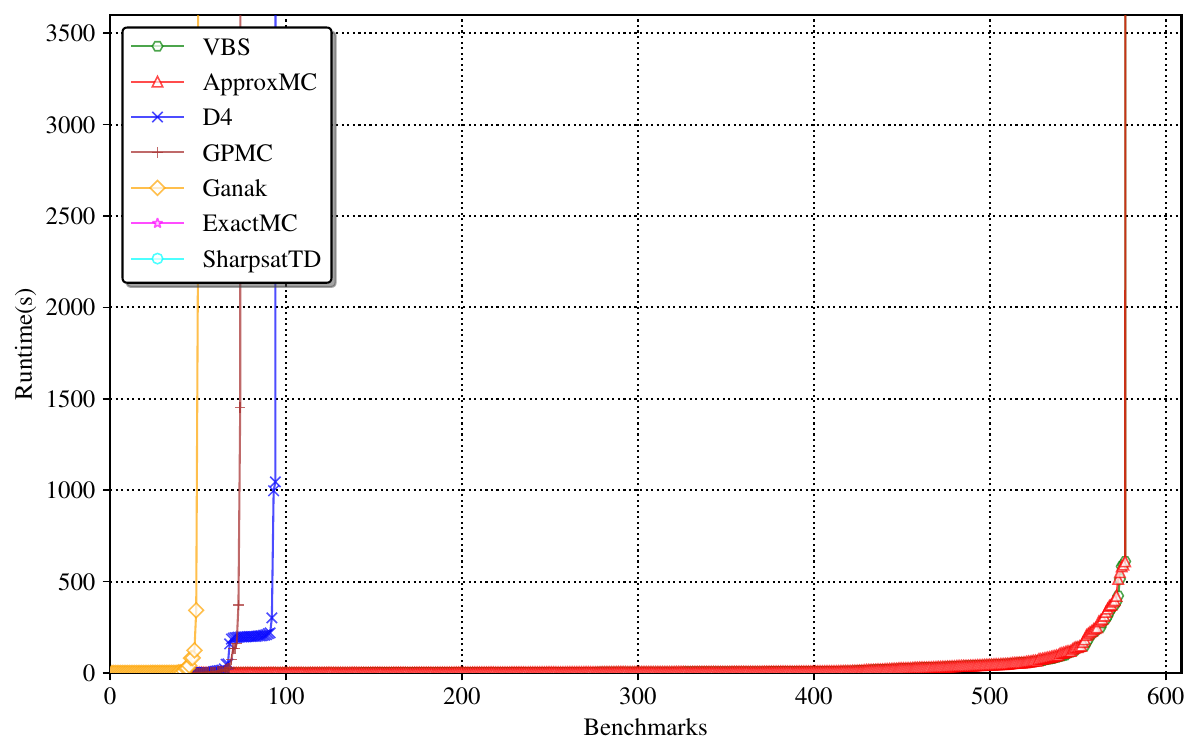}} %
  \end{center}
  \caption{Cactus plot of the numbers of benchmarks solved by model counters on different benchmark sets. }
  \label{fig:cactuss-2}
\end{figure*}

\begin{figure*}[t]
  \begin{center}
    \subfloat[ApproxMC]{\includegraphics[width=0.45\linewidth]{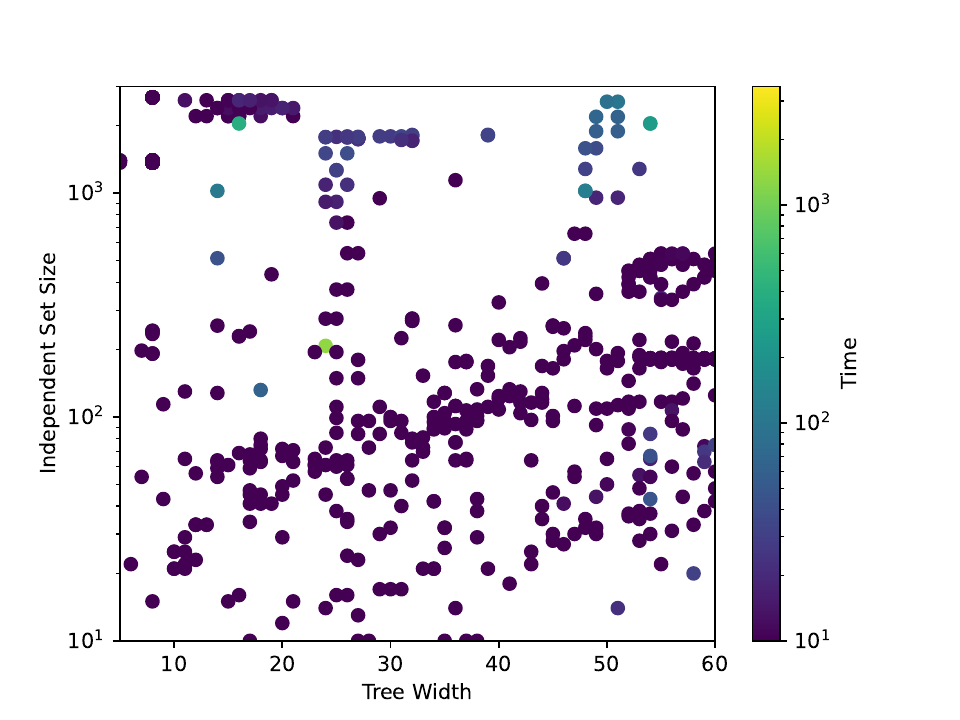}}
    \subfloat[ExactMC]{\includegraphics[width=0.45\linewidth]{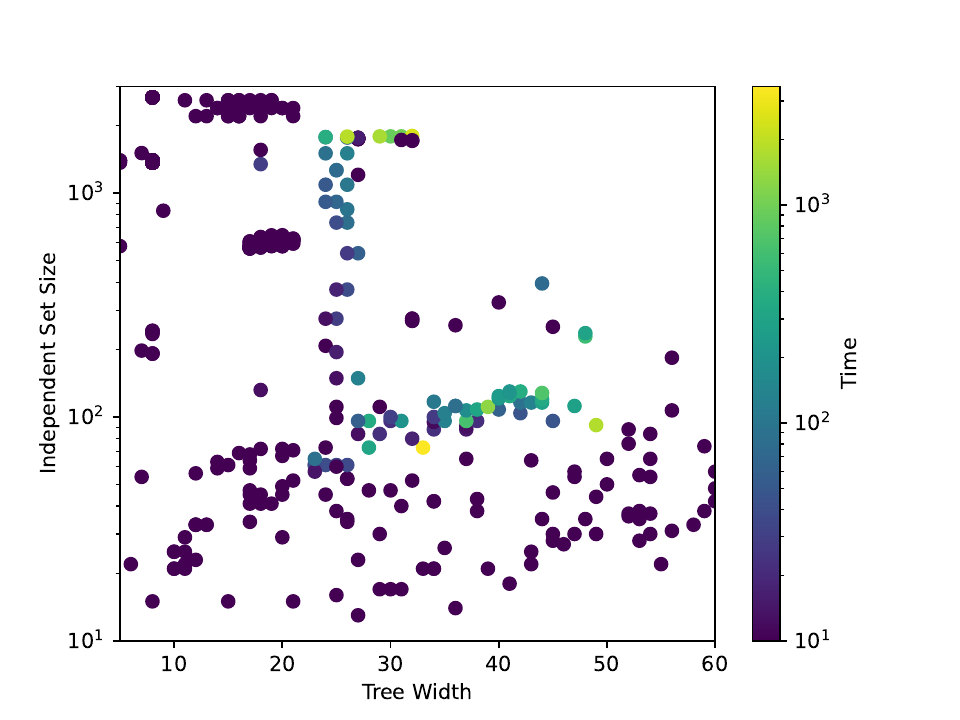}}

  \end{center}
  \caption{Correlation of runtime, independent support size and treewidth (best seen in color).}
  \label{fig:runtime-tw-indep-1}
\end{figure*}

\begin{figure*}[t]
  \begin{center}

    \subfloat[D4]{\includegraphics[width=0.45\linewidth]{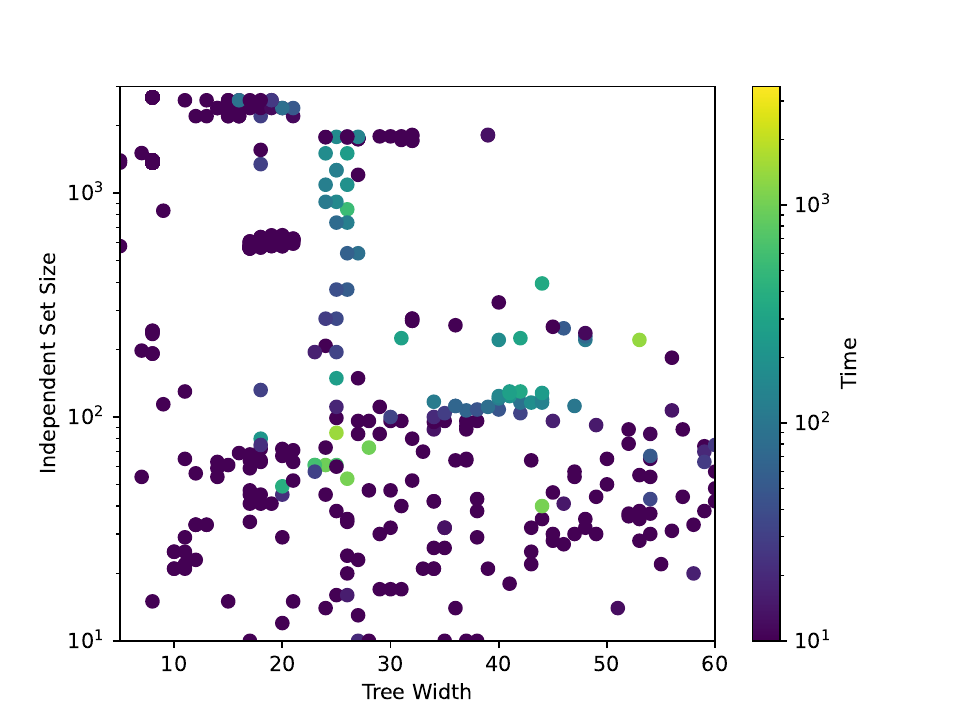}}
    \subfloat[SharpSAT-TD]{\includegraphics[width=0.45\linewidth]{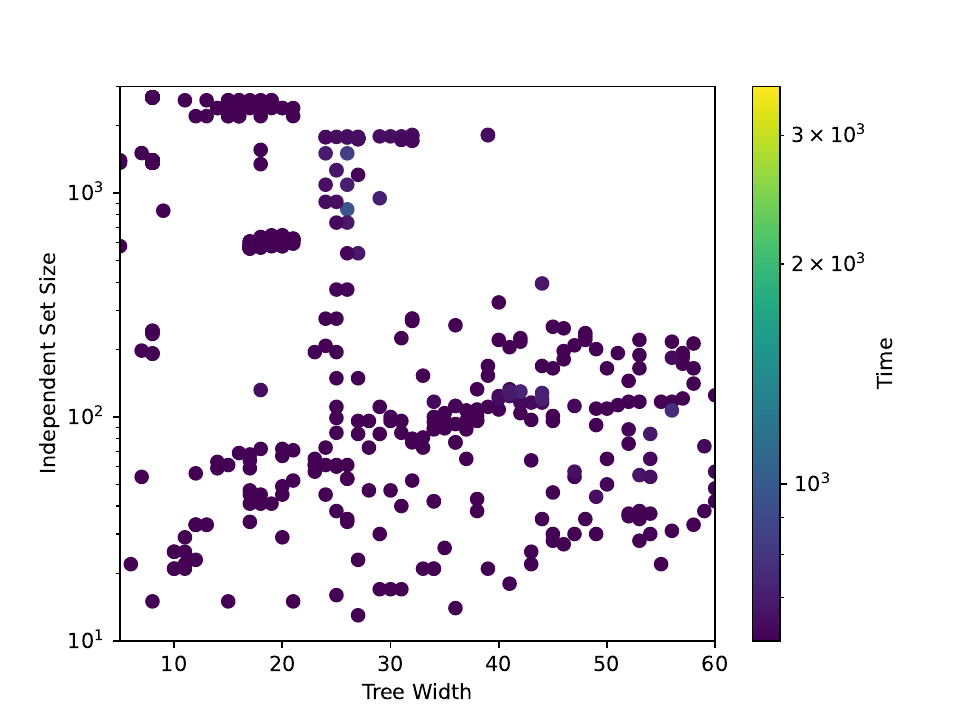}} %
    \\
    \noindent
    \subfloat[GPMC]{\includegraphics[width=0.45\linewidth]{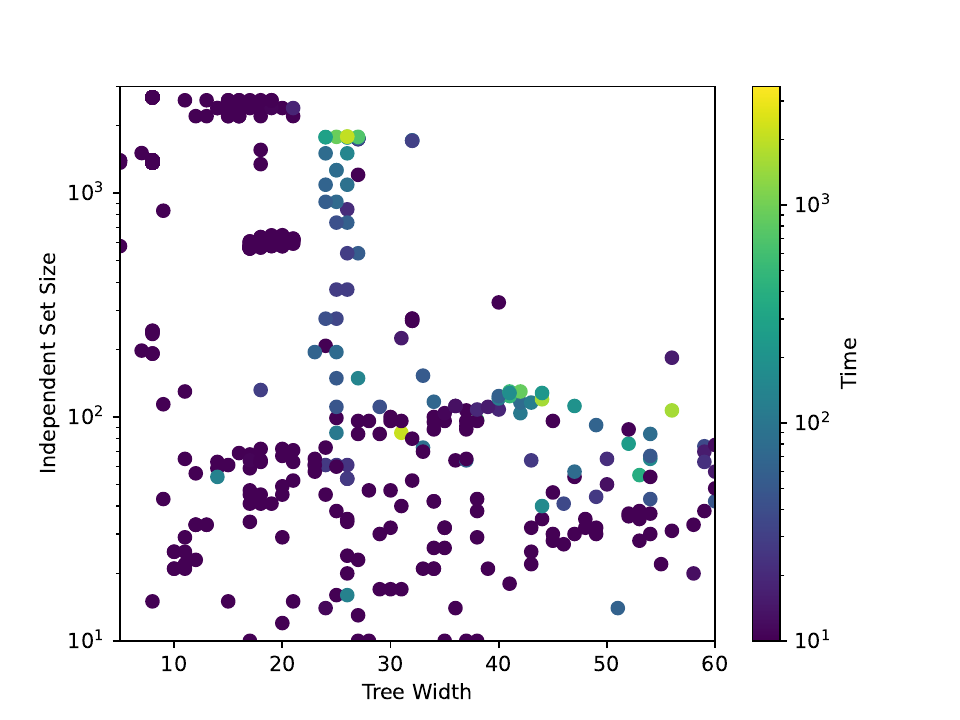}} %
    \subfloat[Ganak]{\includegraphics[width=0.45\linewidth]{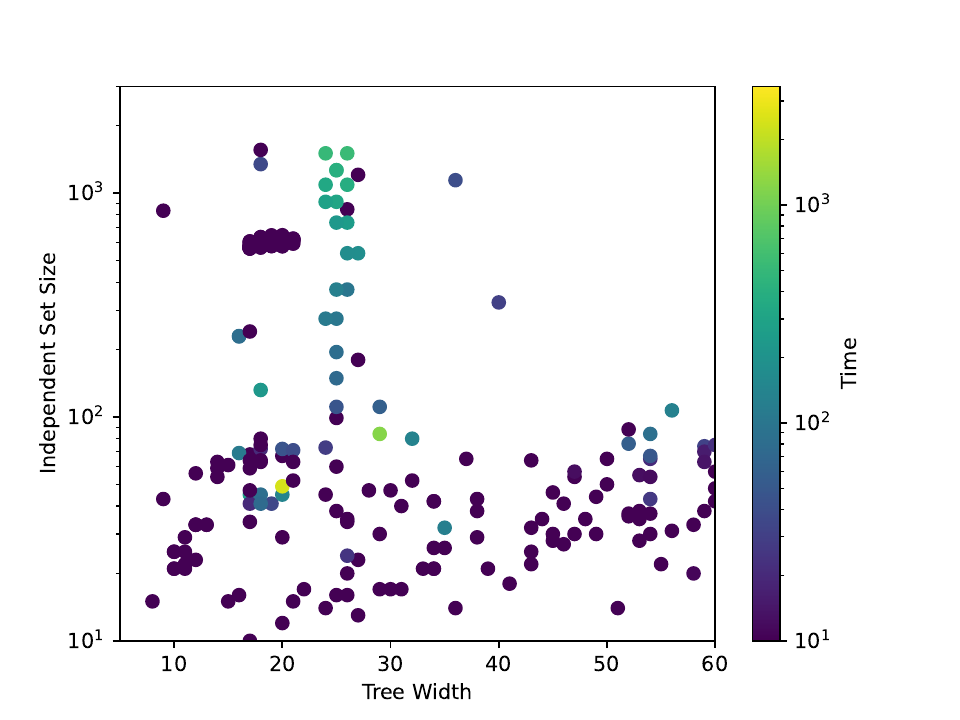}}
  \end{center}
  \caption{Correlation of runtime, independent support size and treewidth (best seen in color).}
  \label{fig:runtime-tw-indep-2}
\end{figure*}

\subsection*{RQ1: Evaluation on Different Benchmark Sets}
We evaluate the performance of the solvers on different benchmark sets by analyzing the total number of instances solved, examining each benchmark set individually, and considering the performance of the virtual best solver.

\subsubsection*{\textsc{Total Benchmarks Solved}}
First, we analyze the total number of instances solved for both non-projected and projected instances.

\mypar{Non-projected Instances.}  \Cref{tab:numsolved-unproj} presents the number of problems solved by various model counters across different benchmark sets. When aggregating all benchmarks, {\sstd} exhibits the best performance, solving 75\% of the benchmarks (811 out of 1080). {\dfour} and {\exmc} also perform well, solving 762 and 745 instances, respectively. In contrast, {\apmc} solves 715 instances, performing relatively less effectively.

\mypar{Projected Instances.} \Cref{tab:numsolved-proj} presents the number of instances solved by various model counters on projected model counting problems. Among the 1182 instances, {\apmc} performs the best, solving 1041 instances. The other projected model counters do not perform as well, with {\gpmc} and {\dfour} solving 434 and 361 instances, respectively, placing them second and third.

\subsubsection{\textsc{Runtime Variation}}
\Cref{tab:numsolved-proj} and \Cref{tab:numsolved-unproj} present the number of problems solved by different model counters across various benchmark sets, highlighting significant performance variations among them. The key observations are as follows:

\begin{enumerate}
  \item Hashing-based counters perform exceptionally well on specific benchmark sets, particularly in cryptographic benchmarks, functional synthesis benchmarks, and neural network verification benchmarks.
  \item Compilation-based counters excel in certain benchmark sets, such as Linux configuration, industrial configuration feature counting benchmarks, and quantitative information flow benchmarks. The performance differences among the search-based counters are minimal.
  \item For the remaining benchmark sets, nearly all counters perform very well.
\end{enumerate}

In \Cref{tab:numsolved-proj} and \Cref{tab:numsolved-unproj}, the dashed lines separate the benchmark sets based on which type of counter performs best.

\Cref{fig:heatmaps} provide a heatmap representation of the percentage of problems each model counter has successfully solved. Each cell in the heatmap indicates the percentage of benchmarks solved by a solver in a specific class of benchmarks. The color scale is shown to the right, with darker colors corresponding to a higher percentage of instances solved.

The cactus plots in \Cref{fig:cactuss} and \Cref{fig:cactuss-2} provide additional insight into performance variations. In these plots, the $x$-axis represents the number of instances, while the $y$-axis indicates the time taken. A point $(i, j)$ on the plot signifies that a solver completed $j$ benchmarks out of the total set in less than or equal to $i$ seconds. The key insights from the figures are as follows:

\begin{enumerate}
  \item In \Cref{fig:cactuss}~(a), cryptographic instances are either solved within seconds or not at all. The VBS closely follows {\apmc}, indicating that {\apmc} typically has minimal runtime in most cases.
  \item A different pattern emerges for neural network verification in \Cref{fig:cactuss}~(b). Here, the counters take a considerable amount of time to solve the instances. The VBS closely follows the curve of {\gpmc} for approximately 700 seconds, suggesting that {\gpmc} can solve around 200 instances more quickly within this timeout. Between 1000 and 3600 seconds, {\apmc} gradually solves around 150 additional instances, a trend not observed in any other benchmark set.
  \item In \Cref{fig:cactuss}~(d), the information flow benchmarks exhibit another interesting behavior, with counters taking varying times between 0 and 2000 seconds to solve instances, managing to solve a maximum of 90 out of 106 instances. However, the VBS solves all the instances within 200 seconds.
\end{enumerate}

The other cactus plots also show similar patterns. We have not included the cactus plots for robust reachability, control improvisation, and Bayes net benchmark sets because, in these sets, all the counters solve the instances within seconds. Similarly, we skipped the industrial and Linux configuration benchmarks since, in these cases, the compilation-based counters solve all the instances within seconds. In contrast, the hashing-based counter solves a negligible number of instances.

\begin{table}[!btp]
  \centering
  \begin{tabular}{lr}
    \toprule
    Counter  & Contribution \\
    \midrule
    {\apmc}  & 1114         \\
    {\exmc}  & 372          \\
    {\gpmc}  & 368          \\
    {\gnk}   & 186          \\
    {\dfour} & 67           \\
    {\sstd}  & 0            \\ \midrule
    Total    & 2106         \\
    \bottomrule
  \end{tabular}
  \caption{Contribution of the solvers to the VBS.}
  \label{tab:contri-to-vbs}
\end{table}

\subsubsection{\textsc{Virtual Best Solver}} The VBS can solve significantly more instances than any individual solver. For the non-projected instances, out of 1080 instances, the VBS can solve 1048 instances. This number is much higher than any individual counter; the best counter for non-projected benchmarks is {\sstd}, which solves 811 instances, representing 77\% of the instances solved by the VBS. In the case of projected instances, the trend continues, with the VBS demonstrating superior performance.

\mypar{Contribution of Counters to the VBS.} In \Cref{tab:contri-to-vbs}, we show the number of instances contributed to the VBS by each model counter. {\apmc} makes the most substantial contribution to the VBS, accounting for 1114 out of 2106 instances. The following most significant contributors are {\exmc} and {\gpmc}, contributing 372 and 368 instances, respectively. Interestingly, {\exmc} only solves non-projected instances. {\sstd} does not contribute any instances to the VBS, likely because it requires a constant 600 seconds to run the tree-decomposition component before starting the actual counting, resulting in longer execution times compared to other counters.

\subsection*{RQ2. Correlation with benchmark parameters}

While the primary observation with model counters on different types of benchmarks was that the performance varies significantly across benchmarks, we sought to identify the underlying parameters from a formula that influences the difficulty of model counting. We consider \textit{treewidth} and size of \textit{independent support set} of a formula as parameters to predict which count would be efficient to count the formula.
Computing the values of each parameter is a computationally hard problem; therefore, we heuristically determine these values. We use $\mathsf{FlowCutter}$~\cite{S17} for computing the treewidth and {\arjun} for calculating the independent support size. In \Cref{tab:tw-ind}, we list the average treewidth and independent support size for each benchmark set. For neural net verification instances, $\mathsf{FlowCutter}$ timed out, which we denote by N.A. in the table.

\begin{table}[!htbp]
  \centering
  \begin{tabular}{lrr}
    \toprule
                & Treewidth & Support Size \\
    \midrule
    Ganak       & 0.38      & -0.22        \\
    D4          & 0.41      & -0.20        \\
    ExactMC     & 0.42      & -0.18        \\
    SharpSAT-TD & 0.48      & -0.12        \\
    GPMC        & 0.47      & -0.09        \\
    ApproxMC    & -0.03     & 0.27         \\
    \bottomrule
  \end{tabular}

  \caption{Correlation between solver runtime and formula features. \\ \centering (Value ranges from -1 to 1, 0 is no linear correlation.)}
  \label{tab:correlation}
\end{table}

In \Cref{tab:correlation}, we present the Pearson correlation between the different benchmark parameters and the solvers' runtimes. The value ranges between -1 and 1, where a greater absolute value indicates a higher correlation. The key observations are as follows:

\begin{enumerate}
  \item The performance of knowledge-compilation-based model counters has a positive correlation with treewidth, while hashing-based counters do not exhibit such a correlation. Among all counters, {\sstd} shows the highest correlation of 0.48 between treewidth and runtime.
  \item The runtime of the hashing-based counter {\apmc} shows a weak positive correlation of 0.27 with independent support size, whereas the compilation-based counters show no correlation.
\end{enumerate}

In \Cref{fig:runtime-tw-indep-1} and \ref{fig:runtime-tw-indep-2}, we represent the relationship among treewidth, independent support size, and the runtime of a specific solver in a heatmap. The $x$-axis represents the treewidth, and the $y$-axis represents the independent support size. Thus, a point with coordinates $(i,j)$ represents an instance with treewidth $i$ and independent support size $j$. The color of the point indicates the runtime for the solver depicted in the graph. The color scale is shown to the right, with darker colors generally indicating shorter runtimes. These heatmaps shed more light on the lack of correlation between the performance of the counters and the benchmark parameters. The heatmaps reveal the following insights:

\begin{enumerate}
  \item In \Cref{fig:runtime-tw-indep-1}~(a), {\apmc} takes approximately $10^3$ seconds to solve instances of treewidth 10, while most of the instances with treewidth between 50 and 60 are solved in less than 100 seconds. The relationship between independent support size and solving time is also not very clear. There are many instances with independent support sizes above 1000 that are solved in less than 10 seconds.
  \item The performance of {\exmc} in \Cref{fig:runtime-tw-indep-1}~(b) is very different. For instances with treewidth higher than 20, it gradually takes an increasing amount of time. Instances with higher treewidth and higher independent support sizes seem to be particularly challenging for {\exmc}. If the treewidth is greater than 50 and the independent support size is greater than 100, the time taken to solve is generally more than 200 seconds.
  \item The results for {\dfour}  in \Cref{fig:runtime-tw-indep-2}~(a) and  {\gpmc}in \Cref{fig:runtime-tw-indep-2}~(c) are not very different from those of {\exmc}.
  \item The behavior of {\sstd}, however, appears a little different in \Cref{fig:runtime-tw-indep-2}~(b). It solves instances with higher treewidth relatively faster, while instances with treewidth greater than 30 and independent support size greater than 100 take more time to solve.
\end{enumerate}

\section{Conclusion}~\label{sec:concl}
We conducted a comprehensive study on a diverse set of benchmarks, revealing that different solvers excel on different subsets. Our findings indicate that the virtual best solver can solve nearly all instances, a feat unattainable by any individual solver alone. This performance of VBS demonstrates that the complementary strategies employed by various solvers enable them to address distinct sets of instances effectively. Consequently, our study underscores the significance of integrating these approaches or developing a portfolio-based solver model.

\paragraph*{Acknowledgment.}
We are thankful to Jiong Yang and Yash Pote for the many useful discussions and grateful to the anonymous reviewers for their constructive comments on improving this paper. We thank  Gabrielle Beck, Guillaume Girol, Samuel Teuber, and Maximilian Zinkus for providing us with different sets of benchmarks. This work was supported in part by the National Research Foundation Singapore under its NRF Fellowship Programme [NRF-NRFFAI1-2019-0004] and Natural Sciences and Engineering Research Council of Canada (NSERC) [RGPIN-2024-05956]. Part of the work was done when Arijit Shaw was a visiting graduate student at the University of Toronto. The computational work for this article was performed on resources of the National Supercomputing Centre, Singapore (\url{https://www.nscc.sg}).

\end{document}